\documentclass[pra,amssymb,amsmath,twocolumn]{revtex4-1}

\usepackage{color,braket}

\newcommand{\be}{\begin{equation}}
\newcommand{\ee}{\end{equation}}

\begin{document}

\title{Contextuality and nonlocality of indistinguishable particles}
\author{Debajyoti Gangopadhyay} \email{humanphysics2014@gmail.com}
\affiliation{Annada College, Vinoba Bhave University, Hazaribag,
  Jharkhand, India} \author{R.  Srikanth}
\email{srik@poornaprajna.org} \affiliation{Poornaprajna Institute of
  Scientific Research, Bengaluru, India}
\begin{abstract} 
Unlike in the case of distinguishable particles, the concept of  entanglement-- not to mention, nonlocality-- remains
debated in case of indistinguishable particles.   Here,  we  show that certain  existing  all-versus-nothing type of proofs  of
contextuality or nonlocality for distinguishable particles, based on a logical contradiction, may be carried
over to indistinguishable particles.
\end{abstract}

\maketitle

\section{Introduction}

Quantum entanglement,  an intriguing feature of  quantum mechanics and
indeed a broad family of nonclassical generalized probability theories
\cite{aravinda2018hierarchical},  refers to  multipartite states  that
can't  be  expressed  as  a  convex  combination  of  product  states.
Nonlocality \cite{epr, bell, chsh}  refers to statistical correlations
among particles of an entangled system, that prevent the assignment of
properties  to  the  particles  even   if  they  are  well  separated.
Entanglement  and  nonlocality are  now recognized as important resources and applied  in various  areas of  quantum information processing,    among   them    quantum   computation,    cryptography,
communication and metrology (cf. \cite{shenoy2017beyond}).

All  the same,  whereas quantum  entanglement and  nonlocality have  a
clear  formulation   in the  case  of  distinguishable   particle  systems
\cite{horodecki2009quantum}, the issue remains  open to some debate in the
case   of  identical   particles. The state space of multiple identical particles lacks the usual tensor product structure, preventing the methods used for studying the usual notions of correlations and entanglement pertaining to distinguishable particles to be carried over. Identical particles do not possess an operational individuality, and enforcing the relevant exchange statistics creates a formal entanglement, whose operational significance is not obvious. The essential problem is that of eliminating the exchange contributions to the entanglement. 

Conventional quantum information theory has normally ignored this issue, since typical practical situations involve distant particles with their wave functions having negligible spatial overlap, making them effectively distinguishable. However, with the advent of quantum computation technologies, which would require ever larger scale integration, the effect of exchange statistics to quantum correlations would no longer be insignificant.

Various authors have responded to this issue over the past two decades. It has been pointed  out \cite{red91} that relinquishing such
`surplus ontology' as  particle labels would clear  up certain aspects
of  the quantum  statistics in  a  way that  is more  faithful to  the
ontology of Fock space.
In \cite{schliemann2001quantum}, entanglement in a pair of fermionic systems in a pure or mixed state is analyzed in terms of Slater rank and Slater number, which are based on the Slater determinant and are in analogy with Schmidt rank and number in the context of two distinguishable particles. In \cite{eckert2002quantum}, this approach is generalized to the multipartite case. Ref. \cite{wiseman2003entanglement} notes the basic contradictions that arise in quantifying the entanglement of identical particles, between the case of whether they are treated as individuals or as mode excitations, and provides an operational approach to resolve the problem.
 Refs.   \cite{gmw02,  ghiman02,   ghiman04} point out that only those bipartite states of identical particles that are not obtainable via  (anti-)symmetrization of product states can be considered as entangled.     
 In \cite{plastino}, the problem of $N$ fermionic pure states is considered, where $N < D$ and $D$ is the single-particle dimension. Since correlations that exist only to guarantee the fermionic antisymemtrization should be discounted as entanglement, multifermionic states of Slater rank  1 are considered separable, and entangled otherwise, in the context of pure states. On this basis, they provide separability criteria for $N$-fermionic pure states in terms of purity or entropy of the single-particle reduced density operator. 
Ref. \cite{benatti2012bipartite} proposes generalizing the concepts of entanglement and separability beyond the usual tensor decomposition of states to the corresponding algebra of observables.  For a  recent review  of state-based or  mode-based approaches  to the entanglement         of        identical         particles,        see
\cite{benatti2014entanglement}.    

Cabello  and   Cunha  \cite{cabcun}
obtained Bell-Kochen-Specker  (BKS) proofs of  contextuality \cite{ks,
  cabello} for  identical particles, by  formulating state-independent
contextuality in the symmetrized or antisymmetrized space of bosons or
fermions. 

The difference between the  non-classicality of properties and that of
identities in quantum mechanics (QM) is that the former is related to non-commutativity of
observables  in  QM,  whereas  the  latter is  connected  to  counting
statistics.  This  indeterminacy   of  identity,  which  has  recently
received  a  lot  of attention  \cite{sancho,krause,dieks,hol,frkr10},
heightens the quantum `mystery' already evident in the uncertainty in,
and entanglement, of the properties.

In this work, we adopt a new approach to correlations among indistinguishable particles, in particular quantum nonlocality and contextuality. In contrast to entanglement among indistinguishable particles, the issue of nonlocality among indistinguishable particles has received very little attention. Our approach revisits the foundations of Bell nonlocality, and makes an attempt to consider particle indistinguishability at that basic level. By establishing nonlocality for indistinguishable particles, we will have of course established entanglement for them. Our main result is that certain types of  inequalities for contextuality or nonlocality among distinguishable particles can be modified in a simple way to yield corresponding inequalities for indistinguishable particles.

The article is arranged as follows. In Section \ref{sec:new}, we explain the new approach adopted to create Bell-Kochen-Specker type inequalities for indistinguishable particles. The method involves adapting all-versus-nothing type inequalities, which are based on a logical contradiction when classical value assignments are made to potential measurement outcomes (cf. \cite{shenoy2019maximally}). A simple instance of contextuality or nonlocality based on such an inequality is discussed in \ref{sec:prop}. This is then adapted to an instance of contextuality among indistinguishable particles in Section \ref{sec:cont}, and to nonlocality among indistinguishable particles in \ref{sec:nonloc}. Finally, we conclude in Section \ref{sec:conclu}.

\section{A new approach \label{sec:new}}

The contradiction obtained  by the Bell-Kochen-Specker  theorems has its origin
in  the fact  that  observables corresponding  to  properties are  not
required  to commute in  QM,  rendering it  impossible  to embed  the
algebra  of  these observables  in  a  commutative  algebra, taken  to
represent the  classical structure  of the putative  hidden variables.

Bell's theorem  tests whether the correlation $P(ab|xy)$ obtained by measuring inputs $x$ and $y$ on two  distinguishable particles, to obtain corresponding outcomes $a$ and $b$, can be expressed as a joint probability distirbution $P(a,b|x,y) = \sum_\lambda p(\lambda)P(a|x,\lambda)P(b|y,\lambda)$, with $\sum_\lambda p(\lambda)=1$, where $\lambda$ parametrizes any local preparation of the two particles (``localism'') and values $a$ and $b$ are assumed to pre-exist according to a probability distribution determined by $\lambda$ (``realism'').  Correlations that violate Bell's inequality lack a local-realist explanation. The distinguishability of the particles is implicitly assumed here. To make this explicit, correlations that satisfy this inequality may be described as embodying two assumptions: (1a) localism of distinguishable particles; (1b) realism of properties possessed by these distinguishable particles.

In the case of indistinguishable particles, extending the above, the natural assumptions to establish nonlocality would be: (2a) localism associated with particles, (2b) realism associated with them, and (2c) distinguishability of the particles. A proof of nonlocality of indistinguishable particles would then contradict a conjunction of  these three assumptions. That is, an inequality based on this model would test the assertion that ``the correlations can be explained by a model that assumes localism and realism and distinguishability''. A violation of the inquality would indicate that one or more of these three assumptions should be given up.

To our knowledge, no such inequality exists in the literature. Here we shall consider a somewhat different model, where assumption (2c) is absorbed into the other two assumptions.  We shall do this by symmetrizing over the inputs, so that $P(ab|xy)$ is replaced by  $P(\overline{a,b|x,y})$. The overline indicates symmetrization over measurements and would correspond to the idea that we are only concerned with the probability associated with a combination of measurement outcomes, and not about which particle produces a specific outcome. 

This modified underlying model would correspond to the twin assumptions of: (3a) localism among non-distinguished particles, and (3b) realism among non-distinguished particles. An inequality based on (3a) and (3b) would test the assertion that ``given particles that are indistinguishable (or simply, not distinguished), the correlations can be explained by a model that assumes localism and realism of the non-distinguished particles''. Thus, a violation this inequality can be interpreted as signifying the nonlocality of indistinguishable particles. In  applying such an inequality to a quantum mechanical system, we implement the symmetrization procedure by not assigning values to measurements on specific particles, but jointly to the symmetrized measurement operator. 

Rather than consider the problem in its full generality, we shall restrict ourselves to the question of adapting existing Bell-type inequalities for distinguishable particles. How an arbitrary bipartite or multipartite Bell-type inequality would be modified by such a symmetrization requirement would depend on the structure of the inequality. However, this process can implemented in a simple way when applied to inequalities that are obtained from logical contradictions-- the so-called all-versus-nothing type inequalities-- and, furthermore, the set of measurements involved possess a symmetry compatible with the symmetrization argument.

After introducing an illustrative example of an all-versus-nothing type contextuality or nonlocality in the following Section, we shall construct the corresponding inequality for identical particles in the subsequent Section.

\section{All-versus-nothing type contextuality / nonlocality among distinguishable particles \label{sec:prop}}

A simple, and historically perhaps the first, instance of an all-versus-nothing type of Bell's theorem was given for the the  GHZ state  of  three distinguishable qubits in the state \cite{ghz}
\begin{equation}
|\Psi\rangle =  \frac{1}{\sqrt{2}} (|\uparrow\uparrow\uparrow\rangle -
|\downarrow\downarrow\downarrow\rangle),
\label{eq:ghz}
\end{equation}
on which  the mutually  commuting three-qubit measurements  $XYY, YXY,
YYX$ and $XXX$ may be performed, where  $X$ and $Y$ refer to the Pauli
operators  along those  directions, and  a tensor  product is  assumed
between   any  two   operators.    The  state   $|\Psi\rangle$  is   a
+1-eigenstate  of  the first  three  three-particle  operators, and  a
$-1$-eigenstate  of the  last.   A realist  (value-definite) model  to
explain quantum indeterminacy of this system requires that there exist
definite value  assignements to  the $X$'s  and $Y$'s  irrespective of
their measurement context.

That no such assignment is possible is seen from the following table:
\begin{equation}
\begin{array}{cccc}
X & Y & Y & \rightarrow +1 \\
Y & X & Y & \rightarrow +1 \\
Y & Y & X & \rightarrow +1 \\
X & X & X & \rightarrow -1
\end{array} 
\label{tab:ghz}
\end{equation}
Any realist assignment of $\pm 1$  to the $X$'s and $Y$'s will yield a
product  of +1 along  the first  three columns  because there  are two
copies of $X$ or $Y$ along each column.  The product of these products
is $+1$, which contradicts the $-1$ obtained above.

However a  contextual-realistic explanation is  possible: for example,
let all $X$'s and  $Y$'s = 1, except $X_1$, which will  be +1 in first
row and  $-1$ in the  last. That in general  a noncontextual-realistic
explanation   is  impossible   is   the  essential   content  of   the
Bell-Kochen-Specker theorem. Any  classical explanation of the quantum
mechanical  experimental  data  must  come  with  a  `twist'  (context
dependence). Assuming that the distinguishable particles are localized
(say \textit{here},  \textit{there} or \textit{yonder}, respectively),
then  the above  contradiction  also implies  the  impossibility of  a
local-realistic explanation of quantum phenomena \cite{mermin}.

By   contrast,  the   following   example  of   the  `Mermin   square'
\cite{mermin},  can't 
be used to demonstrate nonlocality
(but can be so used in case of
contextuality),   because of  the appearance  of joint
operations (in the third row and column below):
\begin{equation}
\begin{array}{ccc}
IX & XI & XX \\
ZI & IZ & ZZ \\
ZX & XZ & YY. \\
\end{array} 
\label{tab:mermin}
\end{equation}
In Eq. (\ref{tab:mermin}), a tensor product is assumed  between each pair of
operators.  The elements  in a given row or  column commute, and their
products, equals $III$  (and thus gives eigenvalue +1  for any state),
except  the  last  column,   where  the  product  operator  yield  the
eigenvalue $-1$.  If the  the two-particle operators are considered to
carry non-contextual (independent  of the row or column  in which they
appear) value-definiteness, then  this implies a contradiction between
quantum mechanics and noncontextual realism.

\section{Contextuality among Indistinguishable particles \label{sec:cont}}

Analogous to the case of nonlocality of indistinguishable particles, contextuality among indistinguishable particles in our approach follows the scheme outlined in Section \ref{sec:new}, except that the concept of non-contextuality (i.e., independence on measurement context) replaces locality. Accordingly, assumptions (3a) and (3b) are replaced by: (4a) non-contextuality among non-distinguished particles, and (4b) realism among non-distinguished particles. 

A violation of an inequality based on these twin assumptions would  demonstrate that any underlying model to explain the correlations between the three identical particles should be contextual, or non-real or both. We will now realize this inequality by extending the logical contradiction in the GHZ  to  indistinguishable  particles.  Suppose  in place  of $|\Psi\rangle$  in  Eq.   (\ref{eq:ghz}),  we  have  the  state of three indistinguishable fermionic particles in the state
\begin{equation}
|\Psi^\prime\rangle                =                \frac{1}{\sqrt{2}}
(|\uparrow\uparrow\uparrow\rangle                                     -
|\downarrow\downarrow\downarrow\rangle)   \otimes   |h,h,h\rangle,
\label{eq:ghz2}
\end{equation}
where 
$|h\rangle$  the single-particle  spatial state  vector of  a particle
that is  \textit{here}.   Note that the spatial specification is 
not necessary in a proof of contextuality, but we have retained it
in anticipation of the following discussion on nonlocality.

In  Eq.  (\ref{tab:ghz}),  the symmetrized  version of
each of the first three operators is the same, which is the sum of all
three (ignoring  the projector in  position space, which is  common to
all three).  The last operator, $XXX$ remains the same.
  
The first  three rows  of Eq. (\ref{tab:ghz})  are `collapsed'  into a
single  expression, so  that  the required  assignment  for the  state
(\ref{eq:ghz2}) is:
\begin{eqnarray}
\overline{YXY} \equiv XYY + YXY + YYX &\rightarrow& +3, \nonumber \\ X
X X &\rightarrow& -1,
\label{eq:ghz1}
\end{eqnarray}
where  a  projector  to  the   support  (assumed  to  be  compact)  of
$|h\rangle\langle  h|$ in  $\left(\mathcal{L}(R^3)\right)^{\otimes 3}$
is  assumed  to  be appended  to  the  two  operators  in the  lhs  of
Eq. (\ref{eq:ghz1}).  

The  product  rule  imposed  on  the hidden  variable  (HV)  state  of
particles becomes another algebraic  rule, combining a sum and product
rule.  In  this case, it is  readily seen that, as  before, a Bell-Kochen-Specker-like
contradiction arises.  The summands in the RHS of the first equation in
Eq. (\ref{eq:ghz1}) can only take  values $\pm
1$. Since  a measurement of  $\overline{YXY}$ must yield +3,  the only
value that can  be assigned to each the summands  in the expresion for
$\overline{YXY}$ is  +1.  Then, the  same reasoning that leads  to the
BKS contradiction in Eq.  (\ref{eq:ghz}) follows here.

Our above method of extending a correlation inequality among distinguishable particles to that between indistinguishable ones by symmetrizing the measurement operators is suitable (as discussed in Section \ref{sec:new}) to all-versus-nothing type inequalities having certain symmetry in the input settings. It may not work in general. A negative example here would the case of the Mermin square. If   the  operators   in  Eq.    (\ref{tab:mermin})  are   adapted  to indistnguishable particles,  whether bosons or fermions,  then we must
symmetrize them. This alters the structure of the Mermin square to the
rectangle, by collapsing the first two columns into one:
\begin{equation}
\begin{array}{ccc}
IX + XI & XX \\
ZI + IZ & ZZ \\
ZX + XZ & YY \\
\end{array},
\label{tab:newmermin}
\end{equation}
However,  the table in Eq. (\ref{tab:newmermin}) lacks  the product  structure that  was exploited  by Eq.
(\ref{tab:mermin})  to set up  a contradiction.   In what  follows, we
will  denote   symmetrization  of  operators  by   an  overline:  thus
$\overline{IX} \equiv IX + XI$, and so on.

For  each   row,  we   find  $(\overline{IX})(XX)   =  \overline{IX}$,
$(\overline{IZ})(ZZ)  =  \overline{IZ}$  and  $(\overline{XZ})(YY)  =
\overline{XZ}$. Thus  a value  assignment consistent with  the product
rule   would  require,   for  example,   that  either   $\overline{IX}
\rightarrow 0$  or $XX  \rightarrow 1$,  and so on  for the  other two
rows.  In each  case, the state-independence of  the BKS contradiction
for the Mermin square is  lost.  Further, the symmetrized operators in
the first column no longer commute, for $[\overline{IX},\overline{IZ}]
= -2i\overline{IY}$, $[\overline{IX},\overline{XZ}] = -2\overline{XY}$
and $[\overline{IZ},\overline{XZ}]  = 2\overline{YZ}$. No  operator in
Eq.  (\ref{tab:newmermin}) appears  in more than one  context, and the
conditions for testing a  contradiction between non-contextual realism
and quantum mechanics  vanish.  

\section{Nonlocality  among Indistinguishable particles  \label{sec:nonloc}}

We shall now adapt the above instance of GHZ-state based proof of contextuality among indistinguishable particles to obtain a proof of nonlocality among three indistinguishable particles.  As noted in Section \ref{sec:new}, our proof of nonlocality would correspond to assumptions (3a) and (3b), which assert the localism and realism among indistinguishable particles. 

We consider  three identical bosonic
particles, localized in three non-overlapping regions, and existing in
the state
\begin{widetext}
\begin{equation}
|\tilde{\Psi}\rangle          \equiv         \frac{1}{\sqrt{2}}
(|\uparrow\uparrow\uparrow\rangle_{123}                               -
|\downarrow\downarrow\downarrow\rangle_{123})                   \otimes
\frac{1}{\sqrt{6}}\left(\sum_P (-1)^p  \ket{\psi_h(x_1),\psi_t(x_2),\psi_y(x_3)}\right),
\label{eq:ghz0}
\end{equation}
\end{widetext}
where  $j  = h,t,y$  indicates  the  position  (being \textit{here},  \textit{there}  or
\textit{yonder}),  $\psi_j(x)$  the  single-particle spatial  density  operator
corresponding  to a  particle being  \textit{here}. The  spatial part  of the
state vector has been antisymmetrized over all permutations $P$ of the
spatial  indices,  which  can   be  explicitly  given  as  the  Slater
determinant of the three tensor producted terms, namely:
\begin{widetext}
\begin{equation}
\ket{\tilde{\psi}} \equiv \frac{1}{\sqrt{6}}(\ket{\psi_h,\psi_t,\psi_y} - \ket{\psi_h\psi_y\psi_t} -
\ket{\psi_t\psi_h\psi_y} + \ket{\psi_t\psi_y\psi_h} -
\ket{\psi_y\psi_t\psi_h} + \ket{\psi_y\psi_h\psi_t}).
\label{eq:tylda}
\end{equation}
\end{widetext}
The antisymmetrization of the spatial component, together with the antisymmetrized spin component in Eq. (\ref{eq:ghz0}), ensures that the overall bosonic state vector is symmetric under particle label exchange.

Let  $\Pi_h$   denote  the  projector   to  the  compact   support  of
$|\psi_h\rangle$, and similarly with $\Pi_y$ and $\Pi_t$. Further, let
$X_h   \equiv  X  \otimes   \Pi_h$.  Thus,   the  operator   $XYY$  on
distinguishable  particles  becomes the  symmetrized  version of  $X_h
\otimes Y_t \otimes Y_y$, given by:
\begin{widetext}
\begin{equation}
\overline{X_h  \otimes Y_t  \otimes  Y_y} =  X_hY_tY_y  + X_hY_yY_t  +
Y_tX_hY_y + Y_tY_yX_h + Y_yY_tX_h + Y_yX_hY_t,
\end{equation}
\end{widetext}
and so on for $Y_hX_tY_y$, $Y_hY_tX_y$ and $X_hX_tX_y$, where a tensor
product  between   the  operators  is   omitted  where  there   is  no
confusion.  The eigenvalues  of  the symmetrized  operators acting  on
$|\tilde{\Psi}\rangle$ are as follows:
\begin{equation}
\begin{array}{cccc}
\overline{X_h ~Y_t ~Y_y} & \rightarrow +1 \\
\overline{Y_h ~X_t ~Y_y} & \rightarrow +1 \\
\overline{Y_h ~Y_t ~X_y} & \rightarrow +1 \\
\overline{X_h ~X_t ~X_y} & \rightarrow -1,
\end{array} 
\label{tab:ghz0}
\end{equation}
which  mirror  Table  (\ref{tab:ghz})  in the  case  of  unsymmetrized
operators acting on the state (\ref{eq:ghz}). Thus a BKS contradiction
follows  precisely   as  in  the  original   case  of  distinguishable
particles.  

Further,  let $\Pi^+_j  \equiv I_2  \otimes \Pi_j$  ($j=h,t,y$), where
$I_2$ is  the identity  operation on the  spin part.  These projectors
have   orthogonal    support.   It    follows   from  
Eqs. (\ref{eq:ghz0}) and (\ref{eq:tylda}) that
\begin{equation}
\langle\tilde{\Psi}|\overline{\Pi^+_h                         \Pi_t^+
\Pi^+_y}|\tilde{\Psi}\rangle = 1
\label{eq:gmwcond}
\end{equation}
which can be interpreted as asserting that precisely one particle (though
one can't say which one) is \textit{here}, precisely one particle is \textit{there}, and
precisely  one  particle  \textit{yonder},  which is  to  say  that  particle
positions are simultaneously objective \cite{gmw02,ghiman02,ghiman04}.
One  can thus  talk of  the proximal  (\textit{here}) particle,  the distal
(\textit{there})  particle  and  the  other  (\textit{yonder})  particle.   These
identities are  not ontological, i.e.,  they are not  the haecceities.
Instead, they  are spatial  identities that are  phenomenological, and
account for  exactly three particles.   The data of  Eq.  (\ref{tab:ghz0})
can be  viewed as  a contradiction  between local realism of
spin properties of these three effectively geographically-individuated particles.

\section{Discussion and conclusions \label{sec:conclu}}

The  nonlocality of identical particles, as against the weaker resource of entanglement, has been hardly studied, whilst in fact, conceptual issues even in the study of entanglement among identical particles remain debated. The present paper explores a specific method to obtain contextuality or nonlocality inequalities for identical particles.
In the characterization of entanglement among identical particles, two distinct approaches for both bosons and fermions may be discerned: one based on the particle aspect and employing first quantization (for example, Refs. \cite{gmw02,  ghiman02,   ghiman04}), and the other based on the mode aspect and employing second quantization (for example,  \cite{schliemann2001quantum,eckert2002quantum,plastino,benatti2012bipartite,benatti2014entanglement}).

Our work is closer in spirit to the first approach, but uses a novel, axiomatic model in the manner of the derivation of the Bell \cite{bell} or CHSH \cite{chsh} inequalities.  Here, a symmetrization over single particles is used to capture the notion of indistinguishability in the underlying model. 

As our main result, we  showed that certain existing logical-contradiction based proofs of
contextuality  or nonlocality  for  distinguishable  particles can  be
adapted  to indistinguishable  particles. The physical interpretation of  this
result  is  that  any  objective value  assignment  of  properties  to
particles,  which remain  non-distinguished  even  at the  ontological
level, produces  a contradiction. 

A future direction would be to study the possibility of constructing more general inequalities for  nonlocality or contextuality of  indistinguisable particles to be derived from those for distinguishable particles, in particular, inequalities that are not of the all-versus-nothing type.

RS thanks  DST-SERB, Govt.  of  India (project EMR/2016/004019) and AMEF, Bengaluru, for financial  support provided.

\end{document}